\documentclass[fleqn,twoside]{article}
\usepackage{espcrc2}

\usepackage{graphicx}
\usepackage[figuresright]{rotating}

\newcommand {\beq}{\begin{equation}}
\newcommand {\eeq}{\end{equation}}
\newcommand {\beqa}{\begin{eqnarray}}
\newcommand {\eeqa}{\end{eqnarray}}
\newcommand {\n}{\nonumber \\}
\newcommand {\tr}{{\rm tr\,}}

\newcommand {\ee}{\mbox{e}}

\newcommand{\AmS}{{\protect\the\textfont2
  A\kern-.1667em\lower.5ex\hbox{M}\kern-.125emS}}

\title{Lattice superstring and noncommutative geometry}

\author{J. Nishimura\address{
High Energy Accelerator Research Organization (KEK)\\
1-1 Oho, Tsukuba 305-0801, Japan}
}
       
\begin{document}

\begin{abstract}
Recent developments in superstring theory 
and noncommutative geometry are deeply related to the idea of 
Eguchi-Kawai reduction in large $N$ gauge theories which dates back to
early 80s. After a general review on this subject
including revived interests in solving planar QCD,
we present some results in the superstring matrix model suggesting
the dynamical generation of 4d space-time
due to the collapse of the eigenvalue distribution.
We then discuss interesting dynamical properties of field theories 
in noncommutative geometry, which have been revealed
by Monte Carlo simulations of twisted reduced models.
We conclude with a comment on the recent proposal for
a lattice construction of supersymmetric gauge theories based on
reduced models.
%
\vspace{1pc}
\end{abstract}

\maketitle

\section{Introduction}

Superstring theory is a unified theory
where matter and gauge particles are both described by
various oscillation modes, while all the interactions including 
gravity are described simply by joining and splitting of strings.
Unification of interactions is not only motivated for aesthetic reasons
but also suggested by precise measurements of the coupling constants 
in the Standard Model.

In fact superstring theory is so far the only quantum theory 
of gravity that is perturbatively well-defined preserving manifest unitarity.
This is in sharp contrast to the situation with the field theoretical
approach treating the metric as a quantum field, where the theory
is perturbatively unrenormalizable. Here one has to study a regularized
theory nonperturbatively and hope to find a nontrivial UV fixed point,
where one can take the continuum limit.
This has been a topic studied over a decade.
There is still a possibility that a sensible continuum limit can be taken,
but the issue of unitarity remains unclear.

The problem with superstring theory, on the other hand,
is that there are too many perturbatively stable vacua
with various space-time dimensionality, gauge group, matter contents and
so on.
This means that non-perturbative effects are crucial for
understanding the `true vacuum', which hopefully describes our world.

At this point let us recall the history of QCD.
Properties of its vacuum such as confinement and chiral symmetry 
breaking as well as the dynamics of low energy excitations
have been understood by a nonperturbative formulation of gauge theory,
namely the lattice gauge theory.
Likewise matrix models, which provide a nonperturbative formulation
of string theory, are expected to give new insights
into the nonperturbative dynamics of string theory.

The particular type of matrix models we will be discussing
appeared in history as large $N$ reduced models in the context 
of solving SU($\infty$) gauge theory \cite{Eguchi:1982nm}.
The matrix model proposed as a nonperturbative 
formulation of superstring theory in ten dimensions \cite{Ishibashi:1996xs}
can be regarded as one of such models.
Among various dynamical issues, we will focus on an exciting possibility,
which has been discussed by many authors
\cite{Aoki:1998vn,Ambjorn:2000dx,NV,Burda:2000mn,%
Ambjorn:2001xs,exact,sign,Nishimura:2001sx,%
KKKMS,Kawai:2002ub,Vernizzi:2002mu,Imai:2003jb}, 
that the SO(10) symmetry of the model is spontaneously broken down
to SO(4) and  the 4d space-time appears {\em dynamically}.

String theory has deep connections to noncommutative geometry,
and large $N$ reduced models naturally incorporate 
this feature \cite{Connes:1997cr,Aoki:1999vr}.
In fact a certain type of reduced models 
\cite{Gonzalez-Arroyo:1982hz}
provides a lattice regularization
of field theories on noncommutative geometry
\cite{Bars:1999av,Ambjorn:1999ts,Profumo:2001hm}.
We will review some Monte Carlo results 
\cite{Bietenholz:2002ch,Bietenholz:2002ev,Ambjorn:2002nj},
which reveal interesting nonperturbative dynamics of such theories.

The reduced model describing nonperturbative superstrings
has manifest supersymmetry even for finite $N$ and
this feature has been utilized recently for a lattice formulation
of supersymmetric gauge theories \cite{Cohen:2003xe}.
There are also revived interests in solving SU($\infty$) gauge theories
with new ideas for treating massless fermions and topologies of
the gauge field \cite{Kiskis:2002gr,Narayanan:2003fc}.
In this review we also cover these new topics
clarifying their mutual relationship.

This article is organized as follows.
In Section \ref{section:matrix}
we discuss the connection between matrix models and string theory.
In Section \ref{section:reduced}
we introduce the large $N$ reduced models
and discuss their equivalence to SU($\infty$) gauge theories.
We also review some recent proposals in this direction.
In Section \ref{section:superstring}
we describe the large $N$ reduced model proposed as 
a nonperturbative formulation of superstring theory
and discuss in particular the dynamical generation of 4d space-time.
In Section \ref{section:noncommutative}
we discuss the connection between large $N$ reduced models 
and noncommutative geometry.
We present Monte Carlo results for Yang-Mills theory and $\phi^4$ theory
on noncommutative spaces,
and discuss their intriguing dynamical properties.
In Section \ref{section:orbifolding} we comment on the 
relation to the recent proposal for a lattice formulation of 
supersymmetric gauge theories.
Section \ref{section:summary} is devoted to a summary and conclusions. 

\section{How matrix models describe strings}
\label{section:matrix}

In this Section we review briefly how matrix models are related to
string theory.
For illustration let us consider a simple matrix model 
defined by the action 
\beq
S_{\rm mat} = \frac{1}{2} N \tr \phi^2 
- \frac{1}{3} N \lambda \, \tr \phi^3 \ ,
\label{onematrix}
\eeq
where $\phi$ is a $N\times N$ Hermitian matrix.
This model can be solved exactly in the large $N$ limit \cite{Brezin:1977sv},
but in order to see the connection to string theory, we
make an expansion with respect to the cubic coupling constant $\lambda$.
Since the dynamical variable $\phi$ has two indices, we use the double-line
notation, which is convenient in discussing the large $N$ limit.

\begin{figure}[htb]
\begin{center}
\includegraphics[height=4cm]{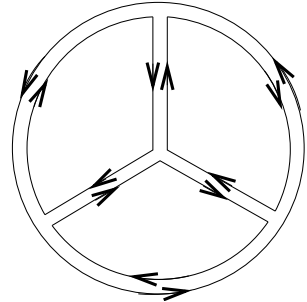}
\caption{
A 3-loop vacuum diagram which appears in the perturbative
expansion of the $\phi^3$ matrix model.
}
\label{fig:benz} 
\end{center}
\end{figure}

A typical vacuum diagram which appears as a O($\lambda^4$) contribution
to the free energy is shown in Fig.\ \ref{fig:benz}.
Each propagator carries $1/N$ and each vertex carries $N \lambda$,
according to the Feynman rule, which can be read off from 
the action (\ref{onematrix}).
A general vacuum diagram ${\cal D}$
with $V$ vertices, $P$ propagators and $I$ index loops
can therefore be evaluated as
\beq
\label{generalD}
{\cal D} \sim \left( \frac{1}{N} \right)^P (N\lambda)^V N^I =
\lambda^V N^\chi \ .
\eeq
An important point here is that 
the power of $N$ is given by the Euler number
$\chi = V - P + I = 2 (1-h)$, which is given by the `genus' $h$,
namely the number of handles of the two-dimensional closed orientable surface,
on which the diagram can be drawn without crossings of lines.
For the diagram in Fig.\ \ref{fig:benz} we have $V=4$, $P=6$, $I=4$, 
hence $h=0$. (Such diagrams are called `planar diagrams' since
they can be drawn on a plane without crossings of lines as in Fig.\
\ref{fig:benz}.)

As one can see from (\ref{generalD}),
if one takes the large $N$ limit for fixed $\lambda$,
only planar diagrams ($h=0$) survive.
Considering the Feynman diagrams as discretized two-dimensional surface,
the continuum limit should be taken along with the fine-tuning of $\lambda$
ensuring that contributions from higher orders in $\lambda$ 
become dominant.
This is known to be possible by sending $\lambda$ to the critical value
$\lambda_{\rm c}$ given by
$(\lambda _{\rm c})^2 = \frac{1}{12 \sqrt{3}}$ \cite{Brezin:1977sv}.
This way of taking the limits (first $N\rightarrow \infty$ and then
$\lambda \nearrow \lambda_{\rm c}$)
is relevant to 2d quantum gravity with
fixed (spherical) topology, if one regards the 2d surface as the space-time.
If instead
one regards the 2d surface as the `worldsheet' 
describing the time-evolution of strings, 
one may view this theory as a classical (or tree-level) string theory,
since there is no branching.

In this regard let us note that the genus $h$ gives
the number of loops in the corresponding diagram in string theory.
In order to obtain all the loop diagrams in string theory,
one has to make diagrams with any genus survive the large $N$ limit.
In the present model this is achieved by 
the so-called double scaling limit, which is
to take the $N\rightarrow \infty$ limit and the $\lambda 
\nearrow \lambda_{\rm c}$ limit 
simultaneously keeping $N^2 |\lambda - \lambda_{\rm c}|^{5/2}$ fixed
\cite{Brezin:rb,Douglas:1989ve,Gross:1989vs}.
This provides a nonperturbative formulation
of bosonic (non-critical) string theory.
In this way one may regard matrix models as `lattice string theory'.

\section{Large $N$ reduced models}
\label{section:reduced}

The connection between matrix models and string theory that
we reviewed in the previous Section is quite general.
But from now on we will focus on a particular class of matrix models
which is known as large $N$ reduced models.
Historically they appeared in the context of solving 
SU($\infty$) gauge theory.

\subsection{Eguchi-Kawai equivalence}

Let us consider SU($N$) lattice gauge theory 
\footnote{In this article we will be sloppy about SU($N$)
or U($N$) to simplify the description
since they become equivalent in the large $N$ limit.
For instance, the $U_\mu$ in (\ref{EKaction}) should be
{\em special}-unitary, and the symmetry (\ref{U1D}) should be $(Z_{N})^D$
if we really start from SU($N$) lattice gauge theory.}
on a $L^D$ lattice 
($D$-dimensional hypercubic lattice with the linear extent $L$)
with the periodic boundary conditions.
Then in the large $N$ (planar) limit, 
Eguchi and Kawai \cite{Eguchi:1982nm}
suggested that
the model with $L=\infty$ is equivalent to 
the model with $L=1$, namely the one-site model defined by the action
\beq
\label{EKaction}
S_{\rm EK}= - N \beta \sum_{\mu\neq \nu} \tr 
(U_\mu U_\nu U_\mu^\dag U_\nu ^\dag   ) \ .
\eeq 
The dynamical variables in this Eguchi-Kawai model are given
by $D$ unitary matrices of size $N$.
(This implies a huge reduction of dynamical degrees of freedom
in the large $N$ limit as indicated in the title of Ref.\ 
\cite{Eguchi:1982nm}.)

In the proof of this statement, there was an important assumption that
the U(1)$^D$ symmetry 
\beq
U_\mu \mapsto \ee ^{i \alpha_\mu} U_\mu 
\label{U1D}
\eeq
of the model (\ref{EKaction})
is {\em} not spontaneously broken.
However it was soon noticed by Bhanot, Heller and Neuberger 
\cite{Bhanot:1982sh}
that the U(1)$^D$ symmetry is actually 
broken at large $\beta$ (weak coupling)
for $D > 2$.
This was shown by considering the eigenvalue distribution 
of the unitary matrix $U_\mu$. 
At $\beta < \beta_{\rm c}$ the eigenvalues lie uniformly on
the unit circle, but
at $\beta > \beta_{\rm c}$ the eigenvalues are concentrated around
some point on the unit circle.
Since the U(1)$^D$ transformation (\ref{U1D})
amounts to rotating all the eigenvalues around the unit circle, 
any non-uniform eigenvalue distribution signals the SSB.
This does not occur at $D=2$ in accord with 
the Mermin-Wagner theorem.

\subsection{Remedies to the original proposal}
\label{section:remedies}

After the discovery of the U(1)$^D$ SSB,
remedies to the original proposal have be suggested.
The authors of Ref.\ \cite{Bhanot:1982sh} proposed
the quenched Eguchi-Kawai model.
Their idea was to constrain the eigenvalues of $U_\mu$ to be 
uniformly distributed on the unit circle.
This can be achieved by inserting 
\beq
\int d V_\mu \, \delta (U_\mu - V_\mu Q V_\mu ^\dag)
\eeq
in the partition function, 
where $Q$ is a diagonal unitary matrix 
$Q = {\rm diag} (1 , \omega , \omega^2 , \cdots , \omega ^{N-1})$
with $\omega = \exp (2\pi i /N)$.

Another proposal was made by Gonzalez-Arroyo and Okawa 
\cite{Gonzalez-Arroyo:1982hz}.
Their idea was to consider a $L^D$ lattice with {\em twisted}
boundary conditions, instead of the periodic ones,
and then set $L=1$. 
The one-site model thus obtained has the action
\beq
\label{TEKaction}
S _{\rm TEK}= - N \beta \sum_{\mu\neq \nu} Z_{\mu\nu} \tr 
(U_\mu U_\nu U_\mu^\dag U_\nu ^\dag   ) \ ,
\eeq
where the Z$_N$ factor $Z_{\mu\nu}$ comes from the twist in the 
boundary conditions. This model is called
the twisted Eguchi-Kawai model.
Modifying the boundary conditions does not alter the thermodynamic limit 
$L=\infty$, but it does change the property of the $L=1$ model drastically.
The configurations which minimize the action are now given by $U_\mu ^{(0)}$
which satisfies
\beq
U_\mu ^{(0)}U_\nu ^{(0)} = Z_{\mu\nu}^*
U_\nu ^{(0)}U_\mu ^{(0)} \ .
\eeq
The solution to this equation is unique up to the symmetry of the
algebra for appropriate choice of the twist.
Most importantly the minimum-action configuration $U_\mu ^{(0)}$ 
has a uniform eigenvalue distribution, which prevents 
the SSB of U(1)$^D$ at large $\beta$.
The Eguchi-Kawai equivalence of the twisted reduced models
is reexamined in Ref.\ \cite{Profumo:2002cm}.

Recently Narayanan and Neuberger \cite{Narayanan:2003fc}
proposed to consider a partially
reduced model with $L>1$ without quenching or twisting.
The key observation is that the Eguchi-Kawai equivalence holds for 
arbitrary $L$  as far as U(1)$^D$ is not spontaneously broken.
The critical $\beta = \beta_{\rm c} (L)$ at which the SSB occurs
depends on $L$, and in particular $\beta_{\rm c} (L) \rightarrow \infty$
as $L\rightarrow \infty$.
Therefore one can always choose $L$ such that the $\beta$ one wants to 
study lies below the critical point $\beta_{\rm c} (L)$.

Various extensions of the Eguchi-Kawai model were considered
in the 80s.
Matter fields in the adjoint and fundamental representations
have been implemented in Refs.\ \cite{Gross:at,Levine:1982uz,Das:1983pm}.
Extensions to non-gauge theories
\cite{Parisi:1982gp,Gonzalez-Arroyo:vx,Eguchi:1982ta}
and to finite temperature
\cite{Klinkhamer:1983dj,Das_Kogut}
are also studied intensively.
For a comprehensive review on these subjects,
we refer the reader to Ref.\ \cite{Das:1984nb}.

\subsection{Revived interests in planar QCD}

In Ref.\ \cite{Kiskis:2002gr} 
Kiskis, Narayanan and Neuberger 
proposed to study planar QCD along the idea of Eguchi-Kawai reduction.
The underlying assumption is of course that QCD ($N_{\rm c}=3$) is actually 
not far from $N_{\rm c}=\infty$ as evidenced in the glueball mass spectrum.
Then the motivation for going to the limit $N_{\rm c}=\infty$ comes from
the fact that the valence quark approximation
\footnote{This is usually called `the quenched approximation',
but we prefer not to use this word to avoid confusion
with the quenching in the reduced model.}
becomes exact as far as the number of flavors $N_{\rm f}$ is kept finite.
Hence all the pathologies observed with this approximation at $N_{\rm c}=3$
should go away in the $N_{\rm c} \rightarrow \infty$ limit, 
and one can make better sense out of Monte Carlo data without having 
to include the dynamical fermions.
This is very attractive given that the full QCD simulation is still costly.

The reason for the exactness of the valence quark approximation
in the large $N_{\rm c}$ limit can be readily seen by comparing 
the fermion loop and the gluon loop.
In the double-line notation the color index loop running
along the gluon loop is replaced by the flavor index loop in the case
of the fermion loop because the fermion is in the fundamental representation
with respect to the color SU($N_{\rm c}$).
This means that the fermion loop is suppressed by
O($N_{\rm f}/N_{\rm c}$) compared with the gluon loop.

If one takes also the $N_{\rm f} \rightarrow \infty$ limit
together with the $N_{\rm c} \rightarrow \infty$ limit with fixed 
$r \equiv N_{\rm f}/N_{\rm c}$, which is referred to as the 
Veneziano limit, the effects of sea quarks survive.
By varying the parameter $r$, one can smoothly interpolate
between the valence QCD and the full QCD.

For the present purpose, one should use the quenched Eguchi-Kawai model 
\cite{Bhanot:1982sh}
or the partially reduced model \cite{Narayanan:2003fc} because
the twisted Eguchi-Kawai model allows only integer $r$ \cite{Das:1983pm}.
The authors of Ref.\ \cite{Kiskis:2002gr}
also suggest to use the overlap Dirac operator \cite{Neuberger:1997fp},
which enables the inclusion of exactly massless fermions in the model.
The problem with the definition for the topological charge mentioned
at the end of Section \ref{cont_ver} is also solved by considering
the overlap Dirac operator.
Correct chiral anomalies have been reproduced from the reduced 
models in Refs.\ \cite{Kikukawa:2002ms,Inagaki:2003uu}.

\subsection{Continuum version of reduced models} 
\label{cont_ver}

Before concluding this Section
we comment on the `continuum version' of large $N$ reduced models, 
which is important because manifest supersymmetry is
implementable. The matrix model describing superstrings in ten dimensions
falls into this category. This version has also been used in a
lattice construction of supersymmetric gauge theories.

Let us recall that the Eguchi-Kawai model has been obtained by considering 
SU($N$) lattice gauge theory and setting the lattice size to $L=1$.
Similarly let us start with continuum SU($N$) gauge theory
and take the zero-volume limit.
The reduced model one obtains in this way is given by the action
\beq
\label{bosonic_action}
S_{\rm b} = - \frac{1}{4 g^2} 
\tr [A_\mu , A_\nu ]^2 \ ,
\eeq 
where $A_\mu$ ($\mu = 1, \cdots , D$) are $N \times N$ Hermitian matrices.

This model has been studied intensively as the bosonic version
of the superstring matrix model.
The finiteness of the partition function 
(for $D>2$ and sufficiently large $N$), 
which is nontrivial because the integration region of Hermitian matrices 
is non-compact, was shown first numerically 
\cite{Krauth:1998yu,Krauth:1998xh}
and proved later \cite{Austing_Wheater}
(see Ref.\ \cite{Austing:2003cz} for a recent analytic work on 
observables).
Ref.\ \cite{Hotta:1998en} studies the large $N$ behavior 
of various correlation functions and 
in particular the SO($D$) symmetry is shown to be unbroken.
This means that the dynamical generation of space-time expected in the 
$D=10$ supersymmetric model does not occur in the bosonic version.
 
In spite of the importance of such Hermitian matrix models in the context
of superstrings and supersymmetric gauge theories,
their equivalence to large $N$ gauge theory is dubious.
Ref.\ \cite{Anagnostopoulos:2001cb} 
studies the VEV of the Wilson loop numerically
in the $D=4$ bosonic case.
The area law holds in a finite regime, but this regime neither extends
nor shrinks in the large $N$ limit.
The situation in the supersymmetric case is expected to be better
\cite{Ambjorn:2000bf}, 
but the achieved $N$ ($N=48$) is not as large as 
in the bosonic case ($N=768$).

The failure of Eguchi-Kawai equivalence in the Hermitian matrix models
may be attributed to the breaking of the U(1)$^D$ symmetry, 
which now reads
\beq
A_\mu \mapsto A_\mu + \alpha _\mu {\bf 1} \ .
\label{cont_U1D}
\eeq
As in the lattice version, one may consider twisting or quenching
to remedy the situation.
It is known that twisting is possible only formally at $N=\infty$ 
\cite{Gonzalez-Arroyo:1983ac}.
Quenching, on the other hand, seems to work at least perturbatively,
but the naive definition of the topological charge 
$Q = \tr F_{\mu\nu} \widetilde{F}_{\mu\nu}$ vanishes 
identically \cite{Gross:at}.

\section{Lattice superstring}
\label{section:superstring}

\subsection{The IKKT model}

In 1996 Ishibashi, Kawai, Kitazawa and Tsuchiya conjectured
that a simple reduced model provides a nonperturbative definition
of type IIB superstring theory in ten dimensions \cite{Ishibashi:1996xs}.
The model, which is now referred to as the IIB matrix model or 
the IKKT model, can be obtained by taking the zero-volume limit
of 10-dimensional SU($N$) super Yang-Mills theory in the continuum.
The action is given explicitly as
\beqa
\label{IIBaction}
S_{\rm IKKT} &=& S_{\rm b} + S_{\rm f}  \ ,   \\
S_{\rm f} &=& 
 -  \frac{1}{2 g^2}  (\Gamma_\mu)_{\alpha\beta}
\tr ( \Psi_\alpha  [A_\mu , \Psi_\beta] ) \ ,
\eeqa
where $S_{\rm b}$ is the bosonic part (\ref{bosonic_action})
and $\Psi_\alpha$ ($\alpha = 1, \cdots , 16$) are $N \times N$
Hermitian fermionic matrices, which transform as Majorana-Weyl
fermions under SO(10) transformation.
The $16\times 16$ matrices $\Gamma _\mu$ are Weyl-projected 
$\gamma$-matrices in the Majorana representation.

There are by now a number of evidences for this conjecture.
We list some of the most crucial ones.
Firstly the action (\ref{IIBaction}) can be regarded as a 
matrix regularization of the worldsheet action of 
the type IIB superstring theory.
Secondly there are solitonic objects in string theory, which are
known as `D-branes', and the IKKT model contains these objects
as classical solutions. Moreover the interaction between D-branes
calculated in the matrix model agrees with the calculations
in string theory. This is remarkable since the interaction includes
gravity. Thirdly there is an attempt to derive 
the string field Hamiltonian
from Schwinger-Dyson equations in the matrix model
\cite{Fukuma:1997en,Aoki:1998bq}.
The derivation is successful albeit 
with the aid of a crude power-counting argument.
This connection, if completed, would provide a direct proof of 
the conjecture.

There are various dynamical issues that should be addressed
in this model.
Some of them have been studied in simplified versions 
\cite{Hotta:1998en,Ambjorn:2000bf,Bialas:2000gf,Anagnostopoulos:2000mn}.
Here we will focus on the possibility that 4d space-time appears
dynamically in the IKKT model.

\subsection{Emergence of 4d space-time}
\label{section:dyn_gen}

Let us first discuss how the space-time is described in this model 
\cite{Aoki:1998vn}.
For that we diagonalize the matrix $A_\mu$ as
\beq
A_\mu = V_\mu X_\mu V_\mu^\dag  \ ,
\eeq
where $X_\mu$ is a real diagonal matrix 
$X_\mu = {\rm diag} (x_{1\mu} , x_{2\mu} , \cdots , x_{N\mu} ) $.
By integrating out the unitary matrices $V_\mu$ first, 
one obtains diagrams like
Fig.\ \ref{fig:benz}, but now each index loop is associated with 
$\vec{x}_{i}= (x_{i1},x_{i2},\cdots , x_{i10} )$,
which may be viewed as describing the embedding of the worldsheet
into the 10-dimensional target space.
Thus we find that the eigenvalues of $A_\mu$ represent the
space-time coordinates.
This means in particular that 
the space-time is treated dynamically in this model.
If the eigenvalue distribution of $A_\mu$
collapses to a 4d hypersurface in the 10d space-time,
we are going to obtain 4d space-time {\em dynamically} in this model.
This requires the SO(10) symmetry of the model to be spontaneously broken.

The order parameter for the SSB can be defined as follows.
Let us define the `moment of inertia tensor' of the space-time as
\beq
T_{\mu\nu} = \frac{1}{N} \tr (A_\mu A_\nu) \ .
\eeq
This matrix has 10 real positive eigenvalues, which we denote
as $\lambda_i$ ($i=1,\cdots , 10$) with the fixed ordering
\beq
\lambda_1 \ge \lambda_2 \ge \cdots \ge \lambda_{10} \ .
\eeq
If the ratio 
$\frac{\langle \lambda_1 \rangle}{\langle \lambda_{10} \rangle}$ 
does not approach unity in the $N\rightarrow \infty$ limit, 
it signals the SSB.
More specifically,
if $\langle \lambda_i \rangle $ with $i=1,\cdots , 4$
turn out to be much larger than
$\langle \lambda_i \rangle $ with $i=5,\cdots , 10$,
it implies the dynamical generation of 4d space-time.

\subsubsection{Gaussian expansion method}

In Refs.\cite{Nishimura:2001sx,KKKMS,Kawai:2002ub}
this issue has been addressed
by the Gaussian expansion method.
In general the method amounts to considering the action 
\beq
S_{\rm GEM}  = \frac{1}{\xi} (S_0 + S  - \xi S_0 )  \ ,
\label{GEMaction}
\eeq
where $S$ is the action of the model one wants to study,
and $S_0$ is a Gaussian action.
If one sets $\xi = 1$ one retrieves the original action $S$.
Then one calculates 
various quantities as an expansion with respect to $\xi$
up to some finite order and set $\xi = 1$.
This yields a loop expansion considering $(S_0 + S)$ as the `classical
action' and $-S_0$ as the `one-loop counter-term'.
The freedom in choosing the Gaussian action $S_0$ is crucial in 
this method.

\begin{figure}[htb]
\includegraphics[width=75mm]{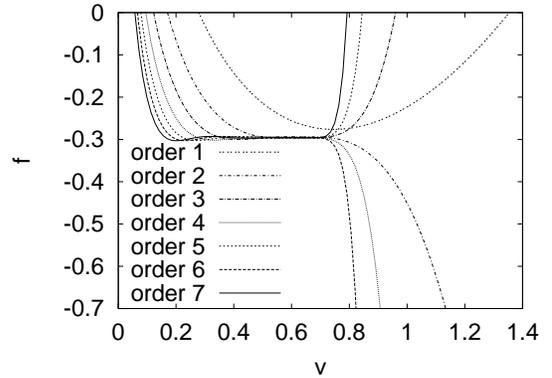}
\caption{
The free energy of the bosonic model 
obtained by the Gaussian expansion method
is plotted as a function of $v$ for $D = 10$.
Each curve corresponds to the order 1, 2, $\cdots$, 7.
}
\label{fig:freeE_10d} 
\end{figure}

\begin{figure}[htb]
\includegraphics[width=75mm]{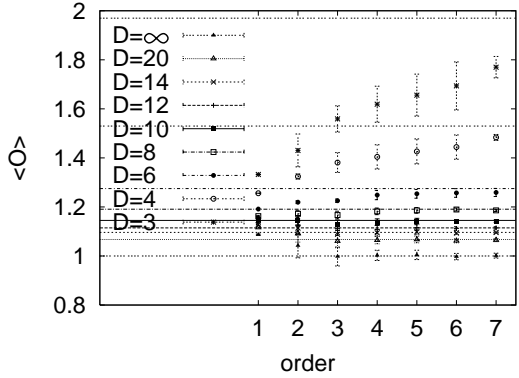}
\caption{
The observable $\langle \frac{1}{N} \tr (A_\mu)^2
\rangle$ 
in the bosonic model
obtained by the Gaussian expansion method
at orders $1,\cdots,7$ for various $D$.
The horizontal lines represent
either
the Monte Carlo results ($D<\infty$) or the exact result ($D=\infty$).
}
\label{fig:extent_10d} 
\end{figure}

Let us first describe how the method works in 
the bosonic version (\ref{bosonic_action}) of 
the IKKT model \cite{Nishimura:2002va},
where Monte Carlo results \cite{Hotta:1998en} are available.
Since we know that the SO($D$) symmetry is not spontaneously broken
in this model \cite{Hotta:1998en}, 
we take the Gaussian action to be
\beq
S_0 = \frac{N}{v} \tr (A_\mu)^2  \ 
\label{Gauss_action}
\eeq
for simplicity,
\footnote{One may also consider SO($D$) breaking Gaussian action
such as $S_0 = \sum_{\mu} \frac{N}{v_\mu} \tr (A_\mu)^2 $.
The method works as well, and reproduces correctly
the absence of SSB in the bosonic model.}
where the parameter $v$ is left arbitrary at this point.

In Fig.\ \ref{fig:freeE_10d} we plot the free energy obtained
in the Gaussian expansion at various orders
as a function  of the parameter $v$ in (\ref{Gauss_action}).
Since $v$ is a parameter which is introduced by hand,
the result should not depend much on it if the expansion becomes valid
for some value of $v$.
Remarkably we do see a clear plateau in Fig.\ \ref{fig:freeE_10d},
and the height of the plateau stabilizes with the increasing order.
Similar behaviors were obtained for
the observable $\langle \frac{1}{N} \tr (A_\mu)^2
\rangle$, and the height of the plateau estimated at each order
(with the `error bar' representing the uncertainty) is shown in 
Fig.\ \ref{fig:extent_10d}. 
The results converge to the horizontal lines,
which represent either
the Monte Carlo results ($D<\infty$) or the exact result ($D=\infty$).
The convergence becomes faster for larger $D$.

\begin{figure}[htb]
\includegraphics[width=75mm]{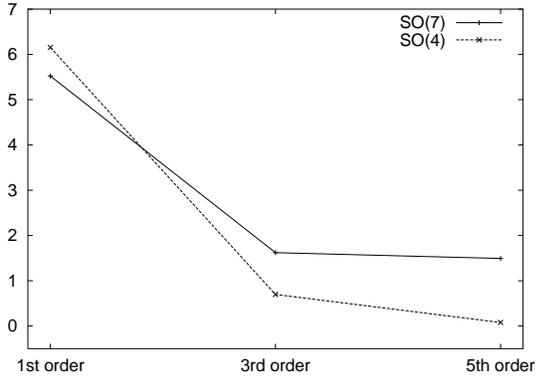}
\caption{
The free energy obtained by the Gaussian expansion method
for the IKKT model at orders 1,3 and 5.
The two symbols correspond to 
the SO(4) Ansatz and the SO(7) Ansatz respectively.
At the 3rd order and higher the free energy for the SO(4) Ansatz
becomes lower than the other.
}
\label{fig:freeE_ikkt} 
\end{figure}

\begin{figure}[htb]
\includegraphics[width=75mm]{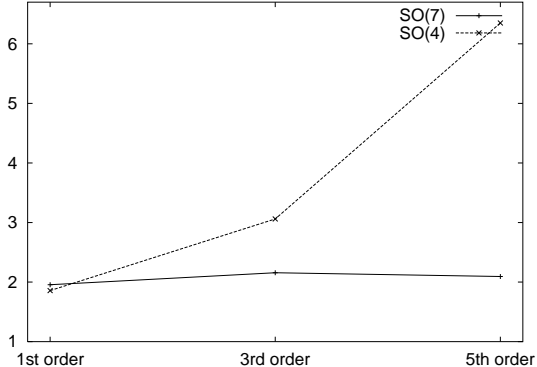}
\caption{
The ratio $R/r$ of the extents 
obtained by the Gaussian expansion method
for the IKKT model at orders 1,3 and 5.
The two symbols correspond to 
the SO(4) Ansatz and the SO(7) Ansatz respectively.
The result for the SO(4) Ansatz grows with the order.
}
\label{fig:ratio_IKKT} 
\end{figure}

Let us turn to the IKKT model (\ref{IIBaction}),
and describe how the first evidence for the 4d space-time was obtained
in Ref.\ \cite{Nishimura:2001sx}.
Since the SSB of the SO(10) symmetry is the main issue here,
the Gaussian action
\beq
S_0 = \sum_{\mu} \frac{N}{v_\mu} \tr (A_\mu)^2  
+  \sum_{\alpha\beta}
N {\cal A} _{\alpha \beta} \tr (\Psi_{\alpha} \Psi_{\beta} )  \ ,
\label{Gauss_action2}
\eeq
which breaks the SO(10) invariance, was considered,
and the Gaussian expansion has been performed up to the 3rd order.
Since it is difficult to search for plateaus 
in the whole space of $v_\mu$ and ${\cal A} _{\alpha \beta}$,
it was assumed that SO($d$) times some
discrete subgroup of SO($10-d$), where $d=2,4,6,7$, is preserved.
This assumption reduced the number of parameters to three,
and plateaus were searched for by finding extrema of the free energy
with respect to the three parameters for each $d$.
At order 3, it was found that the solution for $d=4$ gives 
the smallest free energy.  
The extent of the space-time in the $d$ dimensions ($R$)
and that in the remaining ($10-d$) dimensions ($r$) have been estimated at
the points in the parameter space which extremize the free energy.
The ratio $R/r$ turned out to be larger than one, and for $d=4$
it increased drastically as one goes from order 1 to order 3.

A lot of effort has been made to increase the order of the expansion.
It was soon noticed that 
Schwinger-Dyson equations can be used to reduce
the number of Feynman diagrams to be evaluated \cite{KKKMS}.
Then a computer code 
has been written in order to automatize the task of
listing up and evaluating all the Feynman diagrams \cite{Kawai:2002ub}.
With these technical developments, the order of the expansion
has now been increased up to the 7th order, and the results
strengthened the conclusion of Ref.\ \cite{Nishimura:2001sx}.
In Figs.\ \ref{fig:freeE_ikkt} and \ref{fig:ratio_IKKT} 
we show the free energy and the ratio $R/r$ of the extents
obtained up to the 5th order \cite{KKKMS}.

\subsubsection{Monte Carlo simulation}

Although the results of the Gaussian expansion method are
encouraging and deserve further investigations, it is also
important to confirm the results by Monte Carlo simulation
from first principles. This approach will also allow us
to understand the mechanism for the collapsing of the
eigenvalue distribution of $A_\mu$.

\begin{figure}[htb]
\includegraphics[height=75mm,angle=270]{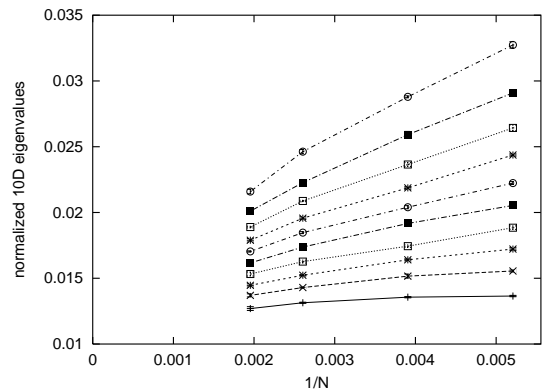}
\caption{
The 10 eigenvalues of the moment of inertia tensor 
in the {\em phase-quenched} model
are plotted against $1/N$.
}
\label{fig:EV_IKKT} 
\end{figure}

An important point to note here is that the fermion determinant
$\det M =  |\det M| \, \ee ^{i \Gamma}$
is actually complex because the fermionic matrices $\Psi_\alpha$
transform as 10d Majorana-Weyl fermion, which is essentially chiral.
(In Euclidean space chiral determinants are generally complex.)
This poses the notorious technical problem known as the 
`complex-action problem'.
As a first step, the phase-quenched model
\beq
\label{phase_quenched}
Z_0 = \int d A \, \ee ^{- S_{\rm b}} |\det M | 
\eeq
has been simulated in Ref.\ \cite{Ambjorn:2000dx}.
We shall denote the VEV with respect to this partition function
by $\langle \ \cdot \ \rangle_0$.
The results for the order parameters $\langle \lambda_i \rangle_0$
are shown in Fig.\ \ref{fig:EV_IKKT}.
If one makes a linear extrapolation to $N=\infty$, one finds that
all the $\langle \lambda_i \rangle_0$ converge to the same value,
meaning that there is no SSB in the phase-quenched model.
This implies that the phase of the fermion determinant plays a crucial
role in the SSB if it happens at all
(Refs.\ \cite{NV,exact} support this scenario).

In order to study the effect of the phase,
a new method has been considered \cite{sign}.
Let us first normalize the eigenvalues of 
the moment of inertia tensor as
\beq
\tilde{\lambda} _i  =
\frac{\lambda _i }{\langle \lambda _i  \rangle _0} \ .
\label{normalize_lambda}
\eeq
Then we define the distribution
\beq
\rho _i (x) = \langle  \delta (x - \tilde{\lambda}_i )  \rangle \  ,
\eeq
which has the factorization property
\beq
\rho _i (x) = \frac{1}{C} \, \rho ^{(0)}_i (x) \, w_i (x) \ ,
\eeq
where $\rho^{(0)} _i (x) = \langle  \delta (x - \tilde{\lambda}_i )
\rangle _0 $ 
denotes the distribution defined in the phase-quenched model 
(\ref{phase_quenched})
and $C=\langle \ee ^{i\Gamma} \rangle_0$ is a normalization constant.

The weight factor $w_i (x)$, which represents 
the effect of the phase,
can be obtained by performing the constrained simulation
\beq
Z_i (x) = \int d A \, \ee ^{- S_{\rm b}} |\det M | 
\, \delta (x - \tilde{\lambda}_i ) \ ,
\eeq
and calculating the expectation value of $\ee ^{i \Gamma}$.
This calculation is the most time-consuming part because of the
oscillating phase,
but one can still do it for moderate $N$.
Various virtues of the method, as compared with the standard reweighting
method,
are discussed in Ref.\ \cite{Ambjorn:2002pz}.

\begin{figure}[htb]
\includegraphics[width=75mm]{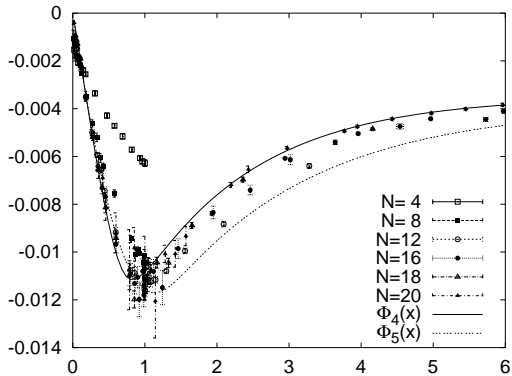}
\caption{
The function $\Phi _4 (x)
\equiv \frac{1}{N^2} \ln w_4 (x) $
is plotted for various $N$.
The solid curve represents a fit to some analytic function.
The dotted curve represents a similar fit to the data for
$\Phi _5 (x)$.
}
\label{fig:Phi_scaling} 
\end{figure}

The distribution $\rho^{(0)} _i (x)$ for the phase-quenched model
is peaked at $x=1$ due to the chosen normalization 
(\ref{normalize_lambda}).
The weight factor $w_i (x)$, on the other hand, turned out 
to have a minimum around $x \sim 1$ 
(see Fig.\ \ref{fig:Phi_scaling}), and as a result 
the distribution $\rho_i(x)$ 
for the full model has a double-peak structure.
Which of the two peaks becomes dominant at large $N$ may depend on $i$.
If it turns out that the peak at $x>1$ dominates for $i\le 4$, but
the peak at $x<1$ dominates for $i\ge 5$, we are able to 
obtain 4d space-time.

A crucial point in this approach is that
\beq
\Phi _i (x) \equiv \frac{1}{N^2} \ln w_i (x)
\eeq
scales with $N$ as one can see from Fig.\ \ref{fig:Phi_scaling}.
This scaling behavior is also understandable by general arguments 
\cite{sign}.
Using the scaling function $\Phi _i (x)$ extracted in this way,
one can estimate the height of the two peaks in 
the distribution $\rho_i(x)$ at much larger $N$.
Preliminary results for $N=64,128$ \cite{sign} are encouraging,
but it remains to be seen whether one can definitely conclude
that 4d space-time appears in the IKKT model.

\section{Lattice noncommutative geometry}
\label{section:noncommutative}

In this Section we discuss noncommutative geometry,
which has deep connections to string theory.
In particular it was shown by Seiberg and Witten 
\cite{Seiberg:1999vs} that
field theories in noncommutative geometry appear
as a certain low-energy limit of string theory, and this
triggered a tremendous boom in this subject.
Among other things it was realized that such theories have
various dynamical properties that ordinary field theories 
do not have. This is due to the UV/IR mixing effect, which
was discovered in the one-loop calculation \cite{Minwalla:1999px}.
Calculations beyond one-loop become complicated because of this effect,
and perturbative renormalizability is not proved even for simple
scalar field theories \cite{Chepelev:1999tt}.
This raises some suspicion that these theories are actually
not well-defined nonperturbatively or the UV/IR mixing effect 
is merely a one-loop artifact.

Fortunately now we know how to regularize these theories on the lattice.
In fact it was found that twisted reduced models at finite $N$
can be interpreted as lattice-regularized field theories in
noncommutative geometry \cite{Ambjorn:1999ts}.
This is a refinement of the earlier work \cite{Aoki:1999vr},
which pointed out the connection between twisted reduced models
and noncommutative field theories.

Noncommutative geometry is characterized by the commutation relation
among the space-time coordinates
\beq
[ \hat{x}_\mu , \hat{x}_\mu ] = i \theta_{\mu\nu} \ ,
\label{NCalg}
\eeq
where $\hat{x}_\mu$ is a Hermitian operator.
When we replace ordinary coordinates $x_\mu$ by the noncommuting
ones $\hat{x}_\mu$, the field $\phi(x)$ should be replaced by
$\hat{\Phi} = \phi(\hat{x})$, which is also an operator.

In fact this setting can be put on a periodic $L^D$ lattice 
in such a way that
there is a one-to-one correspondence between a field $\phi(x)$ on
the lattice and a $N \times N$ matrix $\Phi$.
This means in particular that
\beq
L^D = N^2 
\eeq 
from the matching of the dynamical degrees of freedom.
Using this matrix-field correspondence, one can derive
the {\em noncommutative} lattice field theories
from twisted reduced models \cite{Ambjorn:1999ts}.
Assuming the canonical form for the noncommutative tensor
$\theta_{\mu\nu}$ in (\ref{NCalg}), its scale 
\footnote{In the $D=2$ case, for instance, $\theta$ is
defined by $\theta_{\mu\nu} = \theta \epsilon_{\mu\nu}$.
}
is found to be
\beq
\theta = \frac{1}{\pi} L a^2 \ ,
\eeq
where $a$ is the lattice spacing.

If one takes the planar limit sending $N$ to infinity with fixed $a$,
one obtains $\theta = \infty$.
On the other hand, the twisted reduced model becomes equivalent
to large $N$ field theories in this limit as we discussed in 
Section \ref{section:remedies}.
This is a nonperturbative account of the well-known fact that
noncommutative theories at $\theta = \infty$ are equivalent
to large $N$ theories, which can be easily shown diagrammatically
(see footnote \ref{theta_infty}).

In order to obtain
noncommutative theories at finite $\theta$, one has to take the 
large $N$ limit and the continuum limit simultaneously.
More specifically one has to take the limits $N \rightarrow \infty$ and 
$a\rightarrow 0$ in such a way that $N^{2/D} a^2$ is fixed.
This corresponds to the double scaling limit we discussed in 
Section \ref{section:matrix}.
In what follows we discuss the dynamics of specific
theories, which have been studied by Monte Carlo simulation.

\subsection{NC Yang-Mills theory on the lattice}
\label{section:NCYM}

As the simplest possible gauge theory in
noncommutative geometry,
2d pure Yang-Mills theory has been studied by Monte Carlo simulation
\cite{Bietenholz:2002ch}
using the twisted Eguchi-Kawai model (\ref{TEKaction}).
The planar limit of the theory is solved on the lattice
by Gross and Witten \cite{Gross:he},
so we can refer to the result in our analysis.
Let us consider the VEV of the Wilson loop
$\langle W(I\times I)\rangle$, which is complex in general
unlike in ordinary gauge theory because the noncommutativity $\theta$ 
breaks parity.

\begin{figure}[htbp]
 \begin{center} 
   \vspace{-5mm}
   \includegraphics[width=0.99\linewidth]{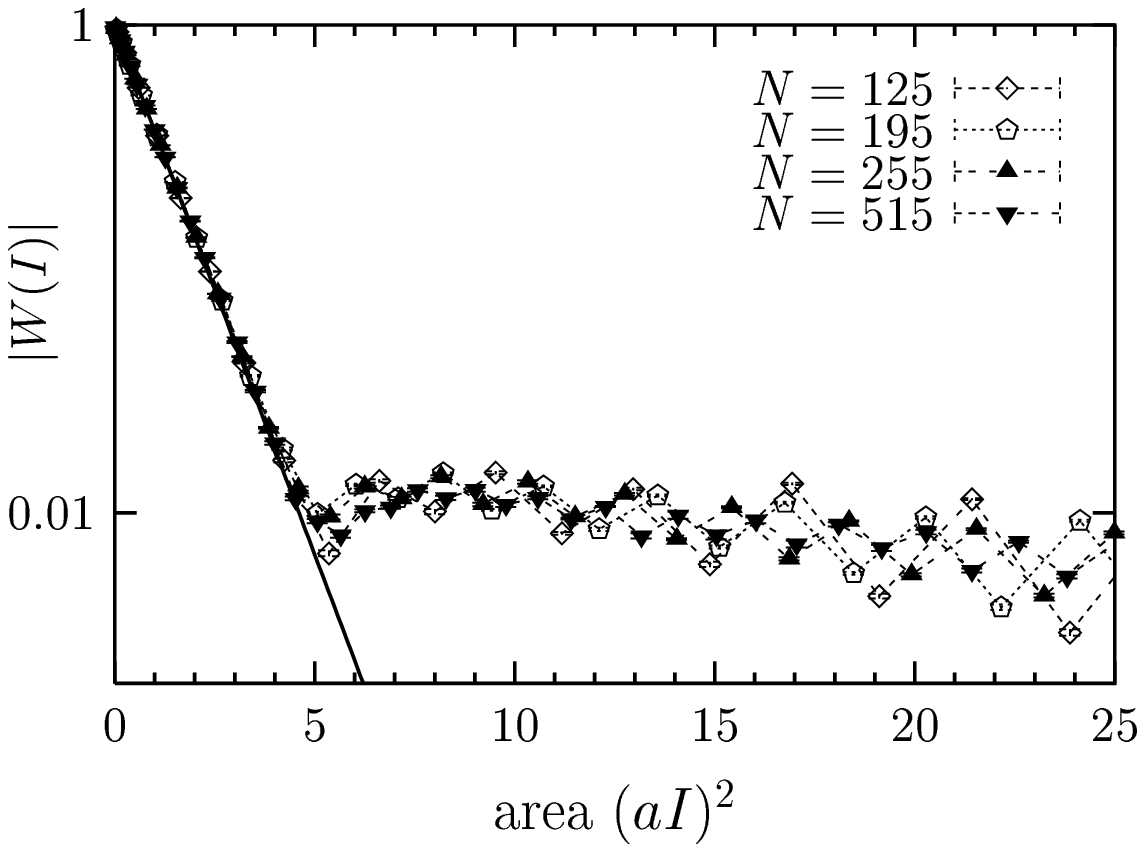}
   \vspace{0.2cm}
   \hspace{.2cm}\includegraphics[width=.95\linewidth]{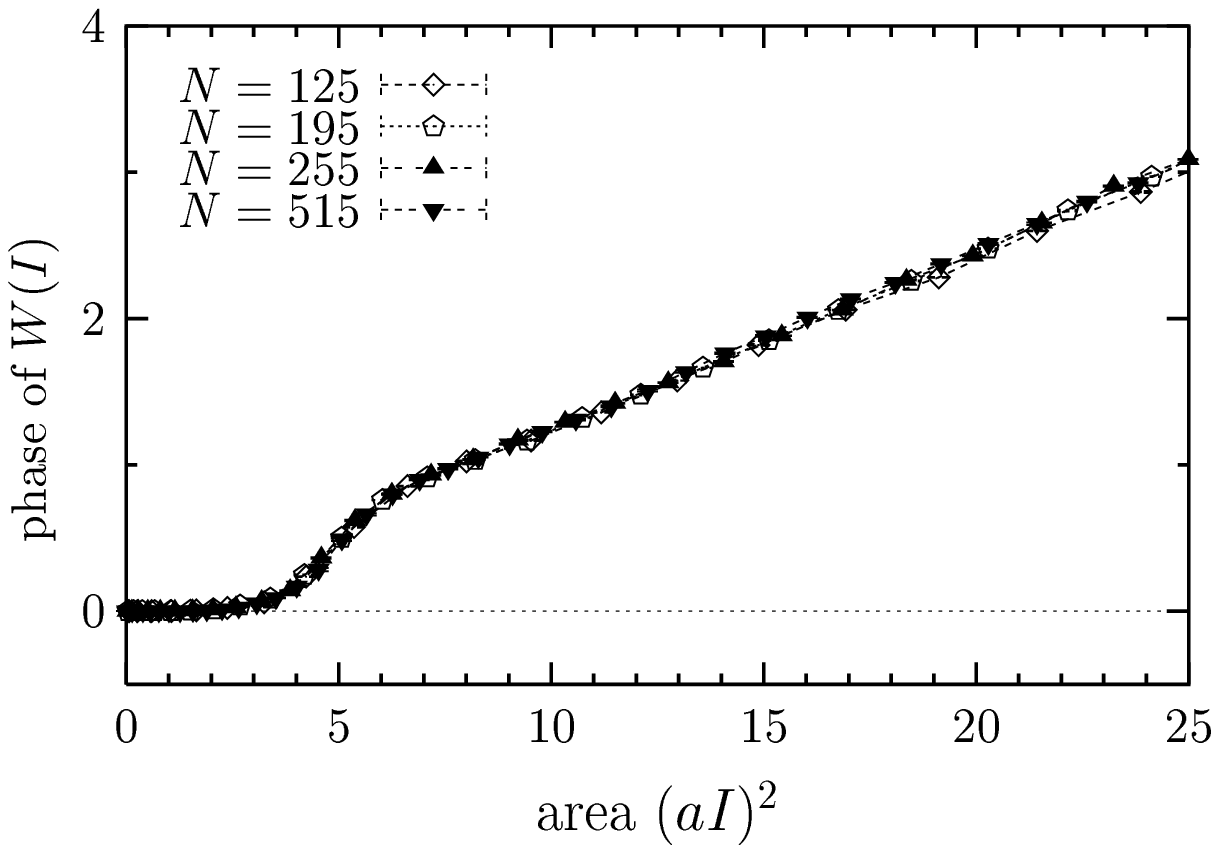}
 \end{center}
\vspace{-10mm}
 \caption{The polar coordinates of the complex Wilson loop $W(I)$
 plotted against the physical area $A = a^2 I^2$.
 At small areas it follows the Gross-Witten
 area law (solid line). 
 At larger areas the absolute value does not decay any more, 
 but the phase increases linearly.}
\label{wil_polar}
\vspace*{-5mm}
\end{figure}


In Fig.\ \ref{wil_polar}
we plot the absolute value and the phase of the Wilson loop
obtained for fixed $\theta = \frac{8}{\pi}$.
One can see that the data in the small area regime $(aI)^2 \ll \theta$
agree with the Gross-Witten planar result.
This is understandable because in the UV regime one cannot see the
finiteness of $\theta$, so the result becomes indistinguishable from
that with $\theta = \infty$, namely the planar result.
The coupling constant $\beta$ has been fine-tuned as a function of the
lattice spacing $a$ in such a way that
the results in the planar regime scale. (In the present case we can use the
Gross-Witten result to infer the tuning of the coupling constant.)
The scaling in this regime is simply a consequence of the fact that
the commutative large $N$ gauge theory has a sensible continuum limit.
What is highly nontrivial is that the scaling extends to larger area,
where nonplanar diagrams start to contribute.
This represents the continuum limit of the noncommutative theory
with finite $\theta$.

In particular we find that 
the phase of the Wilson loop grows linearly with the area as
\beq
({\rm phase}) = \theta ^{-1} \times ({\rm area}) \ ,
\label{ABeffect}
\eeq
which is reminiscent of the Aharonov-Bohm effect if we identify
$\theta^{-1}$ with a static magnetic field traversing the 2d
plane. Such an identification occurs commonly in noncommutative geometry,
but the dynamical AB effect observed here awaits more profound understanding.
Let us also comment that the IR behavior represented by
(\ref{ABeffect}) is clearly different from ordinary gauge theory.
This confirms that the introduction of the noncommutativity $\theta$
can change the IR physics, which is not expected at the classical
level. One should understand this fact as a consequence of the
UV/IR mixing effect caused by the nonplanar diagrams.

\subsection{NC $\phi^4$ theory on the lattice}
\label{section:NCphi4}

Let us discuss the effects of nonplanar diagrams more closely
in the case of scalar field theories.
For the one-loop correction to the inverse propagator, one obtains
\beq
F(p) = \int d^D q \, \frac{\ee^{i \theta_{\mu\nu}q_\nu p_\mu}}
{ q^2 + m^2 } \ ,
\eeq
where the unusual phase factor $\ee^{i \theta_{\mu\nu}q_\nu p_\mu}$
is the only effect of the noncommutative geometry.
\footnote{\label{theta_infty}
In the case of planar diagrams, this `noncommutative phase' cancels.
In the $\theta \rightarrow \infty$ limit, all the nonplanar diagrams
vanish due to the oscillating phase, and only the planar diagrams survive.
This is the diagrammatic account for 
the equivalence of the $\theta = \infty$
theory to the large $N$ field theory.}

\begin{figure}[htb]
\includegraphics[width=75mm]{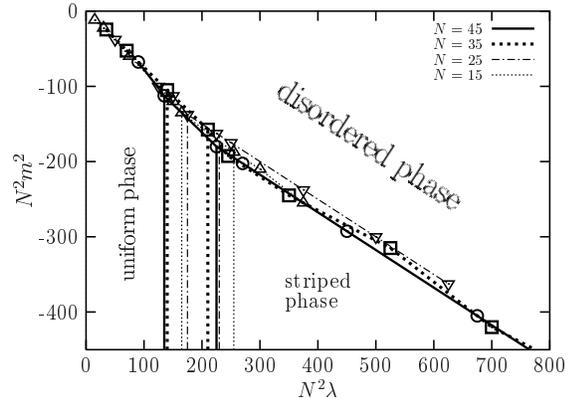}
\caption{
The phase diagram of the $(2+1)$d noncommutative $\phi^4$ theory
on the lattice. 
}
\label{fig:phase_diagram} 
\end{figure}

\begin{figure}[htb]
\includegraphics[width=75mm]{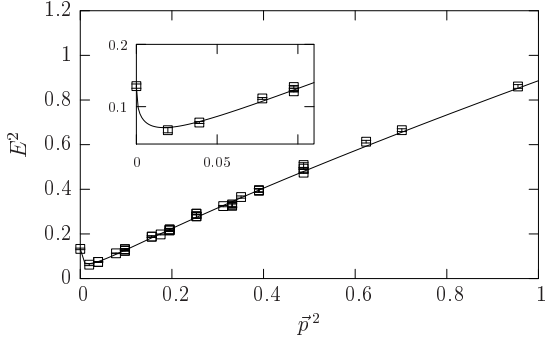}
\caption{
The dispersion relation in the disordered phase
near the critical point.
}
\label{fig:dis_rel} 
\end{figure}


\begin{figure}[htb]
\includegraphics[width=75mm]{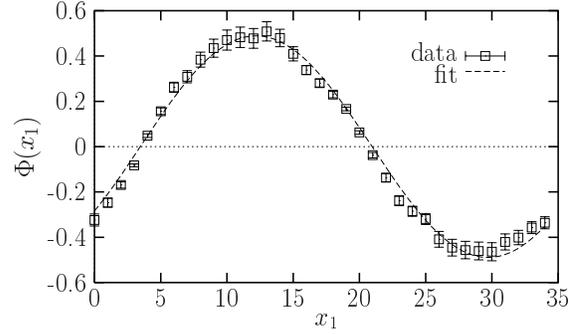}
\caption{
A snapshot of the configuration obtained
just below the critical point with the same $\lambda$ as in 
Fig.\ \ref{fig:dis_rel}.
The average has been taken over the $x_2$ direction, 
in which the value of the scalar field is uniform.
}
\label{fig:snapshot} 
\end{figure}


Because of the `noncommutative phase', the integration over the 
loop momentum $q$ converges, and the function $F(p)$ becomes finite for
$p \neq 0$. At $p=0$ the phase vanishes, and the momentum integration 
has the usual UV divergence. 
This is reflected in the singularity of $F(p)$ at $p\sim 0$ as
\beq
F(p) \sim \frac{1}{|\theta p|^{D-2}} \ .
\label{singularity}
\eeq
Thus the UV divergence of the commutative theory is transformed into
the IR singularity with respect to the external momentum.
This is the famous UV/IR mixing effect \cite{Minwalla:1999px}.

As a physical consequence of this effect, Gubser and Sondhi 
\cite{Gubser:2000cd} conjectured
the existence of the striped phase, where
a nonzero momentum mode condenses, thus giving a position dependent
VEV to the scalar field. This in particular means that the translational
invariance is spontaneously broken.
The analysis was based on self-consistent Hartree-Fock approximation
at the one-loop, and it would be interesting to examine this conjecture
by Monte Carlo simulation.

Let us consider (2+1)d NC $\phi^4$ theory
on the lattice \cite{Bietenholz:2002ev},
where two spatial directions are noncommutative
as in Section \ref{section:NCYM}, 
but now we also have a commuting (Euclidean) time direction
in addition.
The matrix model describing such a theory is given by
\beqa
S_{{\rm NC}\phi^4} &=& N \, \tr \sum_{t=1}^{T}
\left[ \frac{1}{2} \sum _{\mu=1}^{2} 
\left\{ \Gamma_\mu  \Phi(t) \Gamma_\mu ^\dag 
- \Phi(t) \right\} ^2  \right.  \n
&~&   + \frac{1}{2} ( \Phi(t + 1 ) - \Phi(t) )^2  \n
&~& \left. + \frac{m^2}{2} \Phi(t) ^2  
+ \frac{\lambda}{4} \Phi(t) ^4 \right] \ ,
\eeqa
where the shift operators $\Gamma_\mu$ in the noncommuting directions
are defined by 
\beqa
\Gamma_\mu \Gamma_\nu 
&=& Z_{\mu\nu} ^* \Gamma_\nu \Gamma_\mu  \ ,  \\
Z_{\mu\nu} &=& \ee^{\pi i \frac{N+1}{N}} = Z_{\nu\mu} ^*  \ .
\eeqa

Fig.\ \ref{fig:phase_diagram} shows the phase diagram obtained
by Monte Carlo simulation.
The disordered phase appears at large $m^2$ as in the well-known
commutative case.
Let us then decrease $m^2$ for a fixed coupling constant $\lambda$.
When $\lambda$ is small,
the system undergoes the usual Ising-type phase transition
into the uniformly ordered phase.
However, when $\lambda$ is sufficiently large,
one obtains the striped phase as conjectured in Ref.\ \cite{Gubser:2000cd}.
Whether the width of the stripes becomes finite
in the continuum limit
is an interesting open question, which is currently being investigated.

This phenomenon can be understood more clearly by looking at 
the dispersion relation.
Fig.\ \ref{fig:dis_rel} shows the dispersion relation
obtained in the disordered phase near the critical point.
One can see an anomalous behavior at $p \sim 0$, and as a result
the minimum of the energy occurs
at nonzero momentum.
The data can be fitted to 
the form
\beq
E^2 = p^2 + M^2 + \frac{\lambda}{|\theta p|} \ 
\eeq
suggested by the one-loop calculation (\ref{singularity}).
Fig.\ \ref{fig:snapshot} is a snapshot of the configuration
obtained just below the critical point.
The data can be nicely fitted to the sine function,
in accord with the condensation of the nonzero momentum mode.
If $m^2$ is even lower, the shape becomes more step-like.

\section{Lattice supersymmetry via orbifolding}
\label{section:orbifolding}

In this Section we comment on the recent proposal for
a lattice construction of supersymmetric gauge theories
based on reduced models \cite{Cohen:2003xe}.
Let us recall that the continuum version of reduced models
can incorporate manifest SUSY even at finite $N$.
Here the anti-commutator of $Q$ and $\bar{Q}$ yields a generator
of the U(1)$^D$ transformation (\ref{cont_U1D}).
Thus the translation is realized in the internal space,
which is consistent with the identification of the eigenvalues 
of $A_\mu$ as the space-time coordinates described in
Section \ref{section:dyn_gen}.

The key to get a lattice theory from the continuum reduced
model is the `orbifolding', 
which is to impose a constraint on a matrix $\Phi$
such as
\beq
\Omega_j \Phi \Omega_j^\dag = \ee ^{i r_j} \Phi 
\mbox{~~~~~~~~~~~~for~}j=1, \cdots , d \ .
\label{orbifolding}
\eeq
Solving the constraint explicitly, one obtains 
a $d$-dimensional lattice field theory.
In particular the `charge vector' $r_j$ in 
(\ref{orbifolding}) determines that the lattice field corresponding
to the matrix $\Phi$ lives on the links connecting $\vec{x}$ and
$\vec{x}+\vec{r}$.

Part of the SUSY survives the orbifolding, which enables the
restoration of the full SUSY in the continuum limit without fine-tuning.
An extension to the noncommutative geometry is possible 
by making 
$\Omega_j$ in (\ref{orbifolding}) noncommutative 
\cite{Nishimura:2003tf}.

\section{Summary}
\label{section:summary}

We have seen that the idea of Eguchi-Kawai reduction in large $N$
gauge theories has developed in many different directions.
Let us summarize them classifying 
large $N$ reduced models into the lattice version
and the continuum version.

In the lattice version there are a few ways to achieve the equivalence
to large $N$ gauge theories, and there are revived interests in this
direction using the quenching or the partial reduction.
On the other hand, the twisted reduced models have been 
given a new interpretation at {\em finite} $N$
as a lattice regularization of noncommutative field theories.
This motivated a different type of large $N$ limit 
other than the planar limit, where non-planar diagrams also survive.
Monte Carlo studies revealed
intriguing dynamical properties of these theories caused by
the UV/IR mixing.

The continuum version, on the other hand, has been obscure in the 
context of equivalence to large $N$ gauge theories, but
it suddenly became important as a nonperturbative formulation
of superstring theory.
We have discussed that the collapsing of the eigenvalue distribution,
which was problematic in the context of Eguchi-Kawai equivalence,
may provide the key to understand 
the dynamical generation of 4d space-time in 10d superstring theory.
The ultimate goal in this direction
is of course to derive the Standard Model from
the IKKT model or from 
whatever matrix models describing nonperturbative
superstrings. 
%
%




\begin{thebibliography}{9}

\bibitem{Eguchi:1982nm}
T.~Eguchi and H.~Kawai,
Phys.\ Rev.\ Lett.\  {\bf 48} (1982) 1063.

\bibitem{Ishibashi:1996xs}
N.~Ishibashi, H.~Kawai, Y.~Kitazawa and A.~Tsuchiya,
Nucl.\ Phys.\ B {\bf 498} (1997) 467.


\bibitem{Aoki:1998vn}
H.~Aoki, S.~Iso, H.~Kawai, Y.~Kitazawa and T.~Tada,
Prog.\ Theor.\ Phys.\  {\bf 99} (1998) 713.

\bibitem{Ambjorn:2000dx}
J.~Ambj\o rn, K.~N.~Anagnostopoulos, W.~Bietenholz, T.~Hotta and J.~Nishimura,
JHEP {\bf 0007} (2000) 011.


\bibitem{NV}
J.\ Nishimura and G.\ Vernizzi,
JHEP {\bf 0004} (2000) 015; 
Phys. Rev. Lett. {\bf 85} (2000) 4664.


\bibitem{Burda:2000mn}
Z.~Burda, B.~Petersson and J.~Tabaczek,
Nucl.\ Phys.\ B {\bf 602} (2001) 399.

\bibitem{Ambjorn:2001xs}
J.~Ambj\o rn, K.~N.~Anagnostopoulos, W.~Bietenholz, F.~Hofheinz and J.~Nishimura,
Phys.\ Rev.\ D {\bf 65} (2002) 086001.


\bibitem{exact}
J.~Nishimura,
Phys.\ Rev.\ D {\bf 65} (2002) 105012.

\bibitem{sign}
K.~N.~Anagnostopoulos and J.~Nishimura,
Phys.\ Rev.\ D {\bf 66} (2002) 106008.


\bibitem{Nishimura:2001sx}
J.~Nishimura and F.~Sugino,
JHEP {\bf 0205} (2002) 001. 



\bibitem{KKKMS}
H.~Kawai, S.~Kawamoto, T.~Kuroki, T.~Matsuo and S.~Shinohara,
Nucl.\ Phys.\ B {\bf 647} (2002) 153.


\bibitem{Kawai:2002ub}
H.~Kawai, S.~Kawamoto, T.~Kuroki and S.~Shinohara,
Prog.\ Theor.\ Phys.\  {\bf 109} (2003) 115.

\bibitem{Vernizzi:2002mu}
G.~Vernizzi and J.~F.~Wheater,
Phys.\ Rev.\ D {\bf 66} (2002) 085024.



\bibitem{Imai:2003jb}
T.~Imai, Y.~Kitazawa, Y.~Takayama and D.~Tomino,
hep-th/0307007.



\bibitem{Connes:1997cr}
A.~Connes, M.~R.~Douglas and A.~Schwarz,
JHEP {\bf 9802} (1998) 003.

\bibitem{Aoki:1999vr}
H.~Aoki, N.~Ishibashi, S.~Iso, H.~Kawai, Y.~Kitazawa and T.~Tada,
Nucl.\ Phys.\ B {\bf 565} (2000) 176.


\bibitem{Gonzalez-Arroyo:1982hz}
A.~Gonzalez-Arroyo and M.~Okawa,
Phys.\ Rev.\ D {\bf 27} (1983) 2397.


\bibitem{Bars:1999av}
I.~Bars and D.~Minic,
Phys.\ Rev.\ D {\bf 62} (2000) 105018.


\bibitem{Ambjorn:1999ts}
J.~Ambj\o rn, Y.~M.~Makeenko, J.~Nishimura and R.~J.~Szabo,
JHEP {\bf 9911} (1999) 029;
Phys.\ Lett.\ B {\bf 480} (2000) 399;
JHEP {\bf 0005} (2000) 023.

\bibitem{Profumo:2001hm}
S.~Profumo,
JHEP {\bf 0210} (2002) 035.


\bibitem{Bietenholz:2002ch}
W.~Bietenholz, F.~Hofheinz and J.~Nishimura,
JHEP {\bf 0209} (2002) 009.


\bibitem{Bietenholz:2002ev}
W.~Bietenholz, F.~Hofheinz and J.~Nishimura,
Nucl.\ Phys.\ B (Proc.\ Suppl.) {\bf 119} (2003) 941;
%
Fortsch.\ Phys.\  {\bf 51} (2003) 745;
Acta Phys. Polonica B34 (2003) 4711;
F.\ Hofheinz, Ph.\ D.\ Thesis, Humboldt Univ.\ Berlin, 2003;
see also contribution by F.\ Hofheinz to these proceedings (hep-th/0309182).


\bibitem{Ambjorn:2002nj}
J.~Ambj\o rn and S.~Catterall,
Phys.\ Lett.\ B {\bf 549} (2002) 253.



\bibitem{Cohen:2003xe}
A.~G.~Cohen, D.~B.~Kaplan, E.~Katz and M.~Unsal,
JHEP {\bf 0308} (2003) 024;
hep-lat/0307012;
see also contribution by D.B.\ Kaplan 
to these proceedings (hep-lat/0309099).

\bibitem{Kiskis:2002gr}
J.~Kiskis, R.~Narayanan and H.~Neuberger,
Phys.\ Rev.\ D {\bf 66} (2002) 025019.

\bibitem{Narayanan:2003fc}
R.~Narayanan and H.~Neuberger,
hep-lat/0303023.


\bibitem{Brezin:1977sv}
E.~Brezin, C.~Itzykson, G.~Parisi and J.~B.~Zuber,
Commun.\ Math.\ Phys.\  {\bf 59} (1978) 35.

\bibitem{Brezin:rb}
E.~Brezin and V.~A.~Kazakov,
Phys.\ Lett.\ B {\bf 236} (1990) 144.

\bibitem{Douglas:1989ve}
M.~R.~Douglas and S.~H.~Shenker,
Nucl.\ Phys.\ B {\bf 335} (1990) 635.

\bibitem{Gross:1989vs}
D.~J.~Gross and A.~A.~Migdal,
Phys.\ Rev.\ Lett.\  {\bf 64} (1990) 127.



\bibitem{Bhanot:1982sh}
G.~Bhanot, U.~M.~Heller and H.~Neuberger,
Phys.\ Lett.\ B {\bf 113} (1982) 47.

\bibitem{Profumo:2002cm}
S.~Profumo and E.~Vicari,
JHEP {\bf 0205} (2002) 014.


\bibitem{Gross:at}
D.~J.~Gross and Y.~Kitazawa,
Nucl.\ Phys.\ B {\bf 206} (1982) 440.

\bibitem{Levine:1982uz}
H.~Levine and H.~Neuberger,
Phys.\ Lett.\ B {\bf 119} (1982) 183.


\bibitem{Das:1983pm}
S.~R.~Das,
Phys.\ Lett.\ B {\bf 132} (1983) 155.



\bibitem{Parisi:1982gp}
G.~Parisi,
Phys.\ Lett.\ B {\bf 112} (1982) 463.

\bibitem{Gonzalez-Arroyo:vx}
A.~Gonzalez-Arroyo and M.~Okawa,
Nucl.\ Phys.\ B {\bf 247} (1984) 104.


\bibitem{Eguchi:1982ta}
T.~Eguchi and R.~Nakayama,
Phys.\ Lett.\ B {\bf 122} (1983) 59.




\bibitem{Klinkhamer:1983dj}
F.~R.~Klinkhamer and P.~van Baal,
Nucl.\ Phys.\ B {\bf 237} (1984) 274.


\bibitem{Das_Kogut}
S.~R.~Das and J.~B.~Kogut,
Phys.\ Lett.\ B {\bf 141} (1984) 105;
Nucl.\ Phys.\ B {\bf 257} (1985) 141;
Phys.\ Rev.\ D {\bf 31} (1985) 2704.

\bibitem{Das:1984nb}
S.~R.~Das,
Rev.\ Mod.\ Phys.\  {\bf 59} (1987) 235.

\bibitem{Neuberger:1997fp}
H.~Neuberger,
Phys.\ Lett.\ B {\bf 417} (1998) 141.


\bibitem{Kikukawa:2002ms}
Y.~Kikukawa and H.~Suzuki,
JHEP {\bf 0209} (2002) 032.

\bibitem{Inagaki:2003uu}
T.~Inagaki, Y.~Kikukawa and H.~Suzuki,
JHEP {\bf 0305} (2003) 042.

\bibitem{Krauth:1998yu}
W.~Krauth and M.~Staudacher,
Phys.\ Lett.\ B {\bf 435} (1998) 350.

\bibitem{Krauth:1998xh}
W.~Krauth, H.~Nicolai and M.~Staudacher,
Phys.\ Lett.\ B {\bf 431} (1998) 31.


\bibitem{Austing_Wheater}
P.~Austing and J.~F.~Wheater,
JHEP {\bf 0102} (2001) 028;
JHEP {\bf 0104} (2001) 019.


\bibitem{Austing:2003cz}
P.~Austing, G.~Vernizzi and J.~F.~Wheater,
JHEP {\bf 0309} (2003) 023.


\bibitem{Hotta:1998en}
T.~Hotta, J.~Nishimura and A.~Tsuchiya,
Nucl.\ Phys.\ B {\bf 545} (1999) 543.

\bibitem{Anagnostopoulos:2001cb}
K.~N.~Anagnostopoulos, W.~Bietenholz and J.~Nishimura,
Int.\ J.\ Mod.\ Phys.\ C {\bf 13} (2002) 555.

\bibitem{Ambjorn:2000bf}
J.~Ambj\o rn, K.~N.~Anagnostopoulos, W.~Bietenholz, T.~Hotta and J.~Nishimura,
JHEP {\bf 0007} (2000) 013.


\bibitem{Gonzalez-Arroyo:1983ac}
A.~Gonzalez-Arroyo and C.~P.~Korthals Altes,
Phys.\ Lett.\ B {\bf 131} (1983) 396.

\bibitem{Fukuma:1997en}
M.~Fukuma, H.~Kawai, Y.~Kitazawa and A.~Tsuchiya,
Nucl.\ Phys.\ B {\bf 510} (1998) 158.


\bibitem{Aoki:1998bq}
H.~Aoki, S.~Iso, H.~Kawai, Y.~Kitazawa, A.~Tsuchiya and T.~Tada,
Prog.\ Theor.\ Phys.\ Suppl.\  {\bf 134} (1999) 47.



\bibitem{Bialas:2000gf}
P.~Bialas, Z.~Burda, B.~Petersson and J.~Tabaczek,
Nucl.\ Phys.\ B {\bf 592} (2001) 391.


\bibitem{Anagnostopoulos:2000mn}
K.~N.~Anagnostopoulos, J.~Nishimura and P.~Olesen,
JHEP {\bf 0104} (2001) 024.



\bibitem{Nishimura:2002va}
J.~Nishimura, T.~Okubo and F.~Sugino,
JHEP {\bf 0210} (2002) 043.


\bibitem{Ambjorn:2002pz}
J.~Ambj\o rn, K.~N.~Anagnostopoulos, J.~Nishimura and J.~J.~Verbaarschot,
JHEP {\bf 0210} (2002) 062.


\bibitem{Seiberg:1999vs}
N.~Seiberg and E.~Witten,
JHEP {\bf 9909} (1999) 032.


\bibitem{Minwalla:1999px}
S.~Minwalla, M.~Van Raamsdonk and N.~Seiberg,
JHEP {\bf 0002} (2000) 020.

\bibitem{Chepelev:1999tt}
I.~Chepelev and R.~Roiban,
JHEP {\bf 0005} (2000) 037.


\bibitem{Gross:he}
D.~J.~Gross and E.~Witten,
Phys.\ Rev.\ D {\bf 21} (1980) 446.


\bibitem{Gubser:2000cd}
S.~S.~Gubser and S.~L.~Sondhi,
Nucl.\ Phys.\ B {\bf 605} (2001) 395.

\bibitem{Nishimura:2003tf}
J.~Nishimura, S.~J.~Rey and F.~Sugino,
JHEP {\bf 0302} (2003) 032.


\end{thebibliography}
\end{document}